\documentclass[dvips]{article}
\usepackage{4seas95}

\def\lsim{\mathrel{\rlap {\raise.5ex\hbox{$ < $}}
{\lower.5ex\hbox{$\sim$}}}}

\newcommand{\pr}{\paragraph{}}
\newcommand{\be}{\begin{equation}}
\newcommand{\ee}{\end{equation}}
\newcommand{\bea}{\begin{eqnarray}}
\newcommand{\nn}{\nonumber}
\newcommand{\eea}{\end{eqnarray}}

\newcommand{\nk}{\noindent}

\def\gappeq{\mathrel{\rlap {\raise.5ex\hbox{$>$}}
{\lower.5ex\hbox{$\sim$}}}}

\def\lappeq{\mathrel{\rlap{\raise.5ex\hbox{$<$}}
{\lower.5ex\hbox{$\sim$}}}}

\def\beq{\begin{equation}}
\def\eeq{\end{equation}}

\catcode`\@=11 

\def\lsim{\mathrel{\mathpalette\@versim<}}
\def\gsim{\mathrel{\mathpalette\@versim>}}
\def\@versim#1#2{\vcenter{\offinterlineskip
    \ialign{$\m@th#1\hfil##\hfil$\crcr#2\crcr\sim\crcr } }}

\def\t1{{\tilde 1}}

\def\gappeq{\mathrel{\rlap {\raise.5ex\hbox{$>$}}
{\lower.5ex\hbox{$\sim$}}}}

\def\lappeq{\mathrel{\rlap{\raise.5ex\hbox{$<$}}
{\lower.5ex\hbox{$\sim$}}}}

\begin{document}

%
\begin{flushright}
id. no: 95072                \\
preprint no: \\
CTP-TAMU-38-95\\
ACT-13-95. \\
\end{flushright}

\title{ ON A POSSIBLE CONNECTION OF
NON-CRITICAL STRINGS TO CERTAIN ASPECTS OF
QUANTUM BRAIN FUNCTION }

\author{{\bf D.V. Nanopoulos$^{a,\diamond}$}
(speaker)     and
{\bf N. E. Mavromatos$^{b}$}  \\
$^\diamond$ {\it Invited talk at the `four-seas conference',
Trieste (Italy), 25 June-1 July 1995  }  \\
$^{a}$Center for
Theoretical Physics, Dept. of Physics,
Texas A \& M University, College Station, TX 77843-4242, USA
and Astroparticle Physics Group, Houston
Advanced Research Center (HARC), The Mitchell Campus,
Woodlands, TX 77381, USA.  \\
$^{b}$P.P.A.R.C. Advanced Fellow,
Univ. of Oxford, Theoretical Physics,
1 Keble Road, OX1 3NP, U.K. \\
}
\vspace{1cm}
\abstract{
We review certain aspects
of brain function which could be associated with
non-critical (Liouville) string theory. In particular we simulate the
physics of brain microtubules (MT) by using a
(completely integrable) non-critical string, we
discuss the collapse of the wave function as a result
of quantum gravity effects due to abrupt conformational changes
of the MT protein dimers, and we propose a new
mechanism for memory coding.
}

\maketitle
\section{Introduction}
\pr
The interior of living cells is structurally
and dynamically organized by {\it cytoskeletons}, i.e.
networks of protein polymers. Of these structures,
{\it MicroTubules} (MT) appear to be \cite{hameroff}
the most fundamental. Their dynamics has been
studied recently by a number of authors in connection
with the mechanism responsible for
dissipation-free
energy transfer. Recently,
Hameroff and Penrose \cite{HP} have conjectured another
fundamental r${\hat o}$le for the MT, namely being
responsible for {\it quantum computations}
in the human brain, and, thus, related to
consciousness.  The latter is argued to be
associated with certain aspects of quantum theory \cite{penrose}
that are believed to occur in the cytoskeleton MT, in particular
quantum superposition and subsequent collapse of the
wave function of coherent MT networks.
While quantum superposition is a well-established and well-understood
property of quantum physics, the collapse of the  wave function has been
always enigmatic. We review here
one recent proposal~\cite{dn,mn}
to use an explicit string-derived
mechanism -
in one {\it interpretation} of non-critical string theory -
for the collapse of the  wave function\cite{emn},
involving quantum
gravity in an
essential way and solidifying previous intuitively plausible
suggestions\cite{ehns,emohn}.
It is an amazing surprise that quantum gravity effects, of
order of magnitude $G_N^{1/2}m_{p}\sim10^{-19}$, with $G_N$ Newton's
gravitational constant and $m_{p}$ the proton mass,
can play a r\^ole in such low energies as the $eV$ scales
of the typical energy transfer that occurs in cytoskeleta.
However,
the fine details of the MT characteristic
structure
indicate that not only is this conceivable, but such
a picture
leads to order of magnitude estimates for the
time scales entering {\it conscious perception}  that are close enough
to those conjectured/``observed'' by neuroscientists,
based on completely different grounds.
\pr
To understand how quantum space-time effects can affect
conscious perception, we mention that
it has long been suspected \cite{Frohlich}
that large scale quantum coherent phenomena can occur
in the interior of biological cells, as a result of the existence
of ordered water molecules (lattice). Quantum mechanical vibrations
of the latter are
responsible for the appearance of
`phonons' similar in nature to those associated with
superconductivity. In fact there is a close analogy between
superconductivity and energy transfer in biological cells.
In the former
phenomenon electric current is transferred without dissipation
in the surface of the superconductor. In biological cells, as we shall
discuss later on, energy is transferred through the cell
without {\it loss},
despite the existence of frictional forces that represent
the interaction of basic cell constituents
with the surrounding water
molecules \cite{lal}.
Such large scale quantum coherent states can maintain
themselves for up to ${\cal O}(1\,{\rm sec})$,
without significant
environmental entanglement. After that time, the state
undergoes self-collapse, probably
due to quantum
gravity effects. Due to quantum transitions
between the different states
of the quantum system
of MT in certain parts of the human brain,
a sufficient distortion of the surrounding space-time
occurs,
so that a microscopic (Planck size) black hole is formed.
Then collapse is induced, with a collapse time that
depends on the order of magnitude of the number $N$ of
coherent
microtubulins. It is estimated that, with an $N=O[10^{12}]$,
the collapse time
of
${\cal O}(1\,{\rm sec})$,
which appears to be
a typical time scale of conscious events, is achieved.
Taking into account that experiments have shown that
there exist
$N=10^{8}$ tubulins per neuron, and that there are $10^{11}$
neurons in the brain,
it follows that
that this order of magnitude for $N$ refers to a
fraction $10^{-7}$ of the human brain, which is very close to
the fraction believed responsible for human perception.
\pr
The self-collapse of the MT coherent state  wave function
is an essential step for the operation of the MT network
as a quantum computer. In the past it has been suggested
that MT networks processed information in a way similar
to classical cellular automata (CCA)
\cite{hamcel}. These
are described by interacting Ising spin chains on the spatial
plane obtained by fileting open and flattening the MT cylindrical
surface. Distortions in the configurations of individual
parts of the spin chain can be influenced by the environmental spins,
leading to information processing.
In view of the suggestion
\cite{HP} on viewing the conscious parts of the human mind
as quantum computers,
one might extend the concept of the
CCA to a quantum cellular automaton (QCA), undergoing
wave function self-collapses due to (quantum gravity)
enviromental entanglement.
\pr
An interesting and basic
issue that
arises in connection with the above r\^ole of the
brain as a quantum computer
is the emergence of a
direction in the flow of time (arrow). The
latter could be the result
of succesive self-collapses of the system's  wave function.
In a recent series of papers
\cite{emn} we have suggested a rather detailed
mechanism by which an {\it irreversible}
time variable has emerged in certain models of string  quantum
gravity. The model utilized string particles propagating
in singular space-time backgrounds with event horizons.
Consistency of the string approach requires conformal invariance
of the associated $\sigma$-model, which in turn implies a
coupling of the backgrounds for the propagating string modes
to an infinity of
global (quasi-topological) delocalized modes
at higher (massive)
string levels.
The existence of such couplings is necessitated by
specific coherence-preserving target space gauge symmetries
that mix the string levels \cite{emn}.
\pr
The specific model of ref. \cite{emn}
is a completely
integrable string theory, in the sense of being
characterised by an infinity of conserved charges.
This can be intuitively understood by the fact that
the model is a $(1+1)$-dimensional
Liouville string, and as such it can be mapped
to a theory of essentially free fermions
on a discretized world sheet
(matrix model approach \cite{matrix}).
A system of free fermions in $(1+1)$ dimensions
is trivially completely integrable, the infinity of
the conserved charges being provided by
appropriate moments of the fermion energies above the
Fermi surface. Of course, formally, the
precise symmetries of the model used in
ref. \cite{emn}
are much more
complicated \cite{bakas}, but the idea behind the
model's integrability is essentially the above.
It is our belief that
this quantum integrability is a very
important feature of theories of space-time
associated with the time arrow.
In its presence, theories with singular backgrounds
appear consistent as far as maintainance of
quantum coherence is concerned.
This is due to the fact that the phase-space density of the
field theory associated with the matter
degrees of freedom evolves with time according to the
conventional Liouville theorem\cite{emn}
\be
   \partial _t \rho = -\{\rho , H\}_{PB}
\label{one}
\ee
as a consequence of phase-space
volume-preserving symmetries.
In the two-dimensional example of ref. \cite{emn},
these symmetries are known as $W_{\infty}$, and are
associated
with higher spin target-space states\cite{bakas}. They are
responsible for string-level mixing, and hence they are
broken in any low-energy approximation.
If the concept of `measurement
by local scattering experiments' is introduced \cite{emn},
it becomes clear
that the observable
background cannot contain such global modes. The latter have to
be integrated out
in any effective
low-energy theory. The result of this integration
is a non-critical string theory,
based on the propagating modes only.
Its conformal invariance on the world sheet is  restored
by dressing the matter backgrounds by the
Liouville mode $\phi$, which plays the role of the time coordinate.
 The $\phi$ mode is a dynamical local
world-sheet scale \cite{emn},
flowing irreversibly as a result of certain theorems
of the renormalization group of unitary $\sigma$-models
\cite{zam}.
In this way time in target space
has a natural arrow for very specific {\it stringy reasons}.
Eq. (\ref{one}) is now replaced by a modified Liouville
equation with non-Hamiltonian correction terms
\be
   \partial _t \rho = -\{\rho , H\}_{PB}  + \beta ^i G_{ij}
\frac{\partial}{\partial p_j} \rho
\label{modif}
\ee
where $t=ln\phi$ plays the r\^ole of target time, under the
conditions specified in ref. \cite{emn},
$\beta ^i$ is the world-sheet
Renormalization Group $\beta$ function of the
(not exactly marginal)
coupling $g^i$/field mode in target space,
$p_i$ is its
canonical momentum, and
$G_{ij}$ is a `metric'
in the (infinite-dimensional) space of couplings~\cite{zam}.
Upon formulating string theory in higher-genus-resummed
world-sheets, quantization of the couplings/fields $g^i$
is implied. It turns out that canonical quantization
in the space $\{ g^i \}$
is possible in this generalized non-quantum-mechanical
setting~\cite{emninfl}.
\pr
Given the suggestion of ref. \cite{HP}
that space-time environmental entanglement could be
responsible for conscious brain function, it is natural to
examine the conditions under which our theory \cite{emn}
can be applied.
Our approach~\cite{dn,mn}
utilizes extra degrees of freedom, the $W_\infty$
global string modes, which are not directly
accessible to local scattering `experiments' that make
use of {\it propagating} modes only.
Such degrees of freedom carry information, in a similar spirit
to the information loss suggested by Hawking\cite{hawk}
for the quantum-black-hole case. For us, such degrees of freedom
are not exotic, as suggested in ref. \cite{Page},
but
appear {\it already} in the
non-critical String Universe \cite{emn,dn},
and as such they are considered as `purely stringy'.
In this respect, we believe that
the suggested model of consciousness~\cite{dn,mn}, based on the
non-critical-string
formalism of ref. \cite{emn}, is physically more {\it concrete}.
The idea of using string theory instead of
point-like quantum gravity is primarily associated with the
fact that a {\it consistent} quantization of gravity
is at present possible {\it only} within the
framework of string theory, so far. However,
there are additional reasons that make advantageous
a string formalism. These include the possibility of
construction of
a completely integrable model for $MTs$, and
the Hamiltonian
representation of dynamical problems with friction involved in
the physics of MTs. This
leads to
the possibility of a consistent ({\it mean field}) {\it
quantization}
of certain soliton solutions associated with the energy
transfer mechanism in biologcal cells.
\pr
According to our previous discussion
emphasizing the importance of strings,
it is imperative that
we try to
identify the completely integrable
system underlying MT networks.
This will allow
the identification of the
analogue of the (stringy)
propagating degrees of freedom, which
eventually couple to quantum (stringy) gravity
and to global environmental modes.
As we have discussed in detail in ref. \cite{mn}
the relevant basic
building blocks of the human brain are one-dimensional
Ising spin chains,
interacting among themselves in a way so as to
create a large scale quantum coherent state,
believed to be responsible
for preconscious behaviour~\cite{HP}.
The system can be described in a world-sheet
conformal invariant way and is unitary.
Coupling it to gravity generates deviations from conformal
invariance which lead to time-dependences, by identifying
time with the Liouville field (local
renormalization group scale) on the world sheet.
The situation
is similar to the environemtnal entanglement
of ref. \cite{vernon,cald}.
Due to this entanglement,
the system of the propagating modes opens up as in
Markov processes \cite{davidoff}.
This
leads to a dynamical self-collapse of the wave function
of the MT quantum coherent network.
In this way, the part of the human brain associated with
conscioussness generates, through successive collapses,
an arrow of time.
The
interaction among the Ising spin chains,
then, provides a
mechanism for quantum computation,
resembling a planar
cellular automaton.
Such operations sustain the
irreversible flow of time.
\pr
\section{Physical Description of the Microtubules and
their representation as a non-critical string model}
\pr
Microtubules (MT) are hollow cylinders
comprised of an exterior surface
(of cross-section
diameter
$25~nm$)
with 13 arrays
(protofilaments)
of protein
dimers
called tubulins.
The interior of the cylinder
(of cross-section
diameter $14~nm$)
contains ordered water molecules,
which implies the existence
of an electric dipole moment and an electric field.
Ordered water molecules exist also in the exterior
of the MT cylinders, and their presence, together with
the ones in the interior of the MT, is associated
with providing a sort of dissipation. This dissipation
turns out to be quite crucial for an energy-loss-free
trasfer through the MT.
The arrangement of the dimers is such that, if one ignores
their size,
they resemble
triangular lattices on the MT surface. Each dimer
consists of two hydrophobic protein pockets, and
has an electron.
There are two possible positions
of the electron, called $\alpha$ and $\beta$ {\it conformations}.
When the electron is
in the $\beta$-conformation there is a $29^o$ distortion
of the electric dipole moment as compared to the $\alpha $ conformation.
\pr
In standard models for the simulation of the MT dynamics,
the `physical' degree of freedom -
relevant for the description of the energy transfer -
is the projection of the dimer electric dipole moment on the
longitudinal symmetry axis (x-axis) of the MT cylinder.
The $29^o$ distortion of the $\beta$-conformation
leads to a displacement $u_n$ along the $x$-axis,
which is thus the relevant physical degree of freedom.
This way, the effective system is one-dimensional (spatial),
and one has a first indication that quantum integrability
might appear.
\pr
Let $u_n$ be the displacement field of the $n$-th dimer in a MT
chain.
The continuous approximation proves sufficient for the study of
phenomena associated with energy transfer in biological cells,
and this implies that one can make the replacement
\be
  u_n \rightarrow u(x,t)
\label{three}
\ee
with $x$ a spatial coordinate along the longitudinal
symmetry axis of the MT. There is a time variable $t$
due to
fluctuations of the displacements $u(x)$ as a result of the
dipole oscillations in the dimers.
At this stage, $t$ is viewed as a reversible variable.
\pr
The effects of the neighboring
dimers (including neighboring chains)
can be phenomenologically accounted for by an effective
double-well potential of the form suitable for
spontaneous symmetry breakdown \cite{mtmodel}.
The effects of the surrounding water molecules can be
summarized by a viscuous force term that damps out the
dimer oscillations.
This friction should be viewed as an environmental effect, which
however does not lead to energy dissipation, as a result of the
non-trivial
solitonic structure of the
ground-state
and the non-zero constant
force due to the electric field along the MT longitudinal axis.
This is a well known result, directly relevant to
energy transfer in biological systems \cite{lal}.
The solution of the dissipative equations of motion
acquires the form of a travelling wave,
and can
be most easily exhibited by defining a normalized
displacement field
\be
  \psi (\xi ) \propto u (x-vt)
\label{eight}
\ee
with $v$ a propagation velocity determined appropriately
by the parameters of the model~\cite{mtmodel,mn}.
In terms of the $\psi (\xi )$ variable,
the equation of motion of the displacement field
acquires the form of the equation
of motion of an anharmonic oscillator in a frictional environment
\bea
 \psi '' &+& \rho \psi ' - \psi ^3 + \psi + \sigma = 0  \nn \\
\rho &\equiv& \gamma v \sqrt{M |A|(v^2 - v_0^2)^{-\frac{1}{2}}}, \nn \\
\sigma &=& q \sqrt{B}|A|^{-3/2}E
\label{ten}
\eea
with $v_0$ the sound velocity, and
the various other parameters are explained in ref. \cite{mtmodel,mn}.
This has a {\it unique} bounded solution \cite{mtmodel}
\be
    \psi (\xi ) = a + \frac{b -a}{1 + e^{\frac{b-a}{\sqrt{2}}\xi}}
\label{eleven}
\ee
with the parameters $b,a$ and $d$ satisfying:
\be
(\psi -a )(\psi -b )(\psi -d)=\psi ^3 - \psi -
\left(\frac{q \sqrt{B} }{|A|^{3/2}} E\right)
\label{twelve}
\ee
Notice that the kink function (\ref{eleven})
also appears in neural network models, but here
it is derived in a dynamical way.
The kink propagates along the protofilament axis
with fixed velocity
\be
    v=v_0 [1 + \frac{2\gamma}{9d^2Mv_0^2}]^{-\frac{1}{2}}
\label{13}
\ee
with $v_0$ the sound velocity.
This velocity depends on the strength of the electric
field $E$ through the dependence of $d$ on $E$ via (\ref{twelve}).
Notice that, due to friction, $v \ne v_0$, and this is essential
for a non-trivial second derivative term in (\ref{ten}), necessary
for wave propagation.
For realistic biological systems $v \simeq 2 m/sec$.
With a velocity of this order,
the travelling waves
of kink-like excitations of the displacment field
$\psi (\xi )$ transfer energy
along a moderately long microtubule
of length $L =10^{-6} m$ in about
\be
t_T = 5 \times 10^{-7} sec
\label{transfer}
\ee
This time is very close to Frohlich's time for
coherent phonons in biological system.
In realistic biological models for MT,
the effective mass, whose movement might cause distortion
in the surrounding space-time~\cite{HP,dn,mn},
is~\cite{mtmodel} of order
$5 \times 10^{-27} kg$, which is about
the proton mass ($1 GeV$) (!).
As
we discussed in ref. \cite{dn,mn}, these values are essential
in yielding the correct estimates for the
time of collapse of the `{\it preconscious}' state due to our
quantum gravity environmental entanglement.
\pr
To make plausible
a consistent
study of such effects, at a quantum level,
we showed in ref. \cite{mn} that
the equations of motion (\ref{ten})
can be derived from an appropriately constructed
$(1 + 1)$-dimensional (target space) non-critical string model.
Such a representation turns out to be quite crucial,
as it allows for a consistent quantum treatment of
possible gravitational effects, including induced
collapse of the coherent wavefunction of a MT netork~\cite{emn,dn,mn}.
\pr
Abrupt conformational changes of the tubulin dimers, due to
quantum effects (pulses), lead~\cite{mn}  to a formation of {\it virtual}
`black holes' in the effective target two-dimensional
space time. Such black hole solutions can be thought of
either as being associated with real quantum gravitational effects,
or as representing displacement-field-pulse collapse due to abrupt
distortion of the surrounding environment. It should be noted that
in the former case, one can calculate the time required for
an induced collapse of the wave function of a network of MT
consisting of $10^{12}$ dimers. Such quantities
constitute the
part of the brain believed to be responsible for {\it conscious}
perception. The collapse time is found to be
of order ${\cal O}( 1 sec)$ which is in excellent agreement
with observations based on completely different (biological)
methods.
This supports the idea that realistic quantum gravity effects
might play an important r\^ole in conscious  brain functioning.
\pr
The possibility that a (quantum) theory as weak as gravity affects
the physics of low-energy systems, like MT networks,
does not seem so remote, if we recall some recent
{\it experimental} indications \cite{tabony}
about an appreciable sensitivity
exhibited by MT structures to classical gravity effects,
and more general to weak external fields.
In such experiments, gravity effects can lead to
a sort of symmetry breaking and pattern formation
in assemblies of MT. Although theoretically
the sistuation is still very vague, however
the above phenomena appear consistent with
predictions based on reaction-diffucion
theories~\cite{react}, involving
out-of-equilibrium chemical reactions coupled to gravity.
As we have argued in ref. \cite{emn} quantum gravity in
non-critical string theory can be viewed as such
an out-of-equilibrium theory, resulting
in an irreversible flow of time and entropy production
at a fundamental (string) energy scale.
Whether MT systems,
whose (quantum) physics scale is that of electroweak effects,
are senstitive to {\it quantum} gravity effects,
as opposed to classical ones, still remains to be seen.
However the idea does not seem so absurd
if we draw an analogy with what happens
in the neutral kaon system discussed in this meeting~\cite{kaons}.
There, violations of quantum mechanics, associated with
quantum gravity effects of order $O[G_N^{\frac{1}{2}}m_K]
\simeq 10^{-19} $,
could be on the verge of being observed experimentally in $CPLEAR$
or $DA\phi NE$  facilities.
\pr
However we should bear in mind that
in the case of systems pertaining to the
function of the brain things are by no means simple.
The simple fact that the collapse time, calculated on the basis
of string quantum gravity~\cite{dn,mn} or conventional
quantum gravity~\cite{HP} (provided that the latter
exists as a mathematical theory), is in agreement with
estimates of conscious perception time obtained by quite
different methods, although a pleasant indication, however it
by no means constitutes a proof
of the relevance of quantum mechanics or quantum gravity
on brain function. From our point of view~\cite{mn},
such a proof could come from observations of
fluctuations (`quantum jumps') of the length of isolated microtubules.
This is supported by the fact that the Liouville
approach to time in string theory leads~\cite{emn,mn} to
a stochastic growth, a kind of
saw-tooth behaviour of the length of a MT assembly, which notably
is observed experimentally~\cite{sawtooth}.
When applied to an individual MT this approach may lead to
instabilities that could predict fluctuations of the tip
of MT due to quantum gravity entanglement~\cite{rosu,mn}.
\pr
Whether experiments can be devised, which are
sensitive enough to capture such microscopic fluctuations
of isolated (cold) MT, is not known to us.
We believe, however, that if they could be devised, they
would constitute the best proof of the relevance of quantum
(gravity) effects for brain functioning, provided of course that
any other conventional source of mechanical instabilities~\cite{conv}
is excluded. For instance,
the stochastic/diffusive nature
of Liouville gravity, advocated in ref. \cite{mn}, encourages
a comparison with the situation of ref. \cite{tabony} and,
in general, with experiments testing
predictions of reaction-difusion theories.
One is tempted to conjecture that the quantum
fluctuations of the tip of MT structures, predicted in ref. \cite{mn}
in the framework of Liouville theory, might also be seen in the
pattern formation of the experiments of ref. \cite{tabony}, provided
that the latter are repeated in `cold' environment so as to
minimize noise due to
(thermal) dissipation or other mechanical instabilities~\cite{conv},
that could interfere with pure
quantum gravity effects. Whether this is possible, or even conceivable
as an idea for future research, is unknown
to us at present, but we believe that
such speculations deserve closer
attention.

\pr
\section{Memory Coding and Capacity of the brain
as a (non-critical string) dissipative system }
\pr
Irrespective of the possibility of proving experimentally
the possible effects of quantum gravity on brain function,
the conjecture of ref. \cite{mn} that
MT dynamics stems from one-dimensional
Ising spin chains in the brain, that can be represented
as a (completely integrable) non-critical string model
admitting space-time singularities,
implies certain peculiar but highly interesting
properties of brain functioning,
associated with non-equilibrium (dissipative) temporal evolution.
The latter implies an irreversible arrow ot time,
evolution of pure states into mixed ones, and, more generally,
what a particle field theorist would call
$CPT$ violation~\cite{mn}.
\pr
In this respect, our model has many things in common with
dissipative (local field theory) models of ref. \cite{umez,vitiello}
in an attempt to construct realistic models for memory capacity.
Below we shall
briefly review such models, and discuss the possible
advantages
of our (non-critical string)
approach over such local field theory approaches to brain function.
\pr
In conventional brain models~\cite{umez}, based on local field theories,
the kind of symmetry assumed is that of rotational electric dipole
symmetry.
The quantum numbers associated with the latter constitute
a certain class of code numbers.
If the brain lies on a specific ground state, which implies
spontaneous breaking of the dipole rotational symmetry, in order
to reach any other ground state corresponding to a new code number
it would require a sequence
of phase transitions that would destroy the previously stored
information, a procedure known
as {\it overprinting}~\cite{umez,vitiello}.
\pr
A way out of this problem of {\it memory capacity} would be to
increase the symmetry of the problem to the one with
huge dimensions~\cite{stewart}. In a
local field theory this cannot be done
without destroying the practical use of the model.
The problem is analogous to that of how to incorporate
the huge entropy of a macroscopic black hole in an information theory
framework within a local field theory setting. This again would
require an enormous amount of black hole degrees of freedom to
account for the macroscopic entropy, which would be hard, if not
impossible, to reconciliate with the finite number of degrees of
freedom existing in a local field theory.
\pr
String theory seems to provide a way out of these problems~\cite{emn}
due to the infinite-dimensional
gauge stringy symmetries that mix the various levels.
In the black hole model of ref. \cite{witt}, which is used to simulate
the physics of the MT~\cite{mn}, there is an undelrying world-sheet
$SL(2,R)$
symmetry of the $\sigma$-model,
according to which the various stringy states are classified.
The various states of the model, including global string modes
characterised by discrete values of (target) energy and momenta,
are classified by the non-compact isospin $j$ and its
third component $m$, which - unlike the compact
isospin $SU(2)$ case - is not restricted by the  value of $j$.
Thus, for a given $j$, which in the case of string states plays the role
of energy, one can have an {\it infinity} of states labelled by the
value of the third isospin component $m$.
All such states are characterised by a $W_{1+\infty}$ symmetry,
in target space. As we mentioned above, this symmetry
is responsible for the maintenance of
{\it quantum coherence} in the presence of a
black hole background~\cite{emn}, in the sense of an area-preserving
diffeomorphism in a matter phase-space of the two-dimensional target
space theory.
It should be noted that such area-preserving symmetries, as spectrum
generating algebras, also appear in connection with the
excitations of planar quantum Hall systems having non-degenerate
ground states~\cite{trugen}. So, it should not be considered
as a surprise that such symmetries appear in our two-dimensional
spin chain model for the brain MT. In the particular
case of two-dimensional string black holes there is even a
formal analogy with quantum Hall models,
as argued in ref. \cite{emnhall}.
\pr
In ref. \cite{emn} it has been argued that such symmetries
are responsible for an `infinite-dimensional' quantum hair ($W$-hair)
of the two-dimensional black hole, which consists of (conserved)
quantum (global) charges, characterising a black hole space-time
even asymptotically, i.e. after evaporation.
Such a hair would induce a huge degeneracy in the ground state
of the system that could lead to the solution of the problem of
memory capacity. From a formal point of view, the rigorous
existence proof
of such conserved charges would be the
explicit construction of exactly marginal
deformations that correspond to turning on the above charges.
The exaclty marginal character of the deformations is
required in order to maintain
conformal invariance of the world-sheet $\sigma$-model and thus
stable ground state of the string.
\pr
At present, within the black hole model of ref. \cite{witt},
used to simulate also the physics of the MT dynamics~\cite{mn},
it became possible to construct~\cite{chaudh}
the exactly marginal deformation
corresponding
only to the lowest non-trivial charge, which is the
$ADM$ mass of the black hole. From a $W_\infty$ symmetry
point of view, this would be the charge associated with the
spin-two part of the target space spectrum, i.e. the stress-energy
tensor of the black hole.
There is a huge degeneracy of the ground state of the system
which is due to the existence of known exactly marginal deformations
that are responsible for changing continuously the
$ADM$ mass of the black hole.
In the notation of ref. \cite{chaudh}, such deformations
are denoted by $L_0^2 {\overline L}_0^2$, and their coupling constant
(which
is a {\it free} parameter of the model) shifts
the $ADM$ mass of the black hole space time.  The above operator
turns on only backgrounds corresponding to the (discrete)
higher-level string states that do not propagate in space-time.
The ground state of such models consists of turning on
backgrounds corresponding to matter propagating states.
Such backgrounds are turned on by another exactly marginal
deformation $L_0^1 {\overline L}_0^1$, which
mixes
the propagating states (belonging to the lowest string level)
with an infinity of higher-level string global states.
Both operators $L_0^1$ and $L_0^2$ owe their existence to
the target space $W_{1+\infty}$-spectrum-generating algebra
of the black hole space-time~\cite{emn,chaudh}. The latter
is broken explicitly by
`measurement' by
local scattering experiments or in general by operations
that are performed within localised regions of space-time,
such as those taking place in the conscious part of the brain.
\pr
Such a procedure will integrate out the global degrees of freedom,
leaving only an effective (string) theory of propagating degrees
of freedom in a black hole background space time.
For each matter ground state of a propagating degree of freedom,
say the zero mode of the massless field corresponding to the static
``tachyon'' background of ref. \cite{witt}, with $SL(2,R)$
quantum numbers
$j=-\frac{1}{2}, m=0$, there will be an infinite degeneracy
corresponding to a continuum of black hole space-time backgrounds
with different $ADM$ masses.
These backgrounds are
essentially generated by
adding various constants to the configuration of the dilaton
field in this
two-dimensional string theory~\cite{witt,chaudh}.
It should be noted here that the infinity of
propagating
``tachyon'' states (lowest string mass-level (massless) states),
corresponding to other values of $m$, for
continuous representations of $j$, constitute
{\it excitations} about the ground state(s), and, thus,
they should not be considered as contributing to the
ground state degeneracy.
In principle, there may be an additional infinity of
quantum numbers corresponding to higher-level $W$-hair charges
of the black hole space time which are
believed~\cite{emn}
responsible for quantum coherence at the full string theory level.
\pr
Taking into account the conjecture of ref. \cite{mn},
that formation of virtual black holes
can occur in brain MT models, which
would correspond to different modes of collapse of
pulses of the displacement field $\psi $
defined in (\ref{eight})~\cite{mn},
one obtains a system of {\it coding} that is capable to
solve in principle the problem of memory capacity.
Information is stored in the brain in the following sense:
every time there is an external stimulus that brings the brain
out of equilibrium, one can imagine an abrupt conformational
change of the MT dimers, leading to a collapse
of the pulse pertaining to the displacement field. Then a (virtual)
black hole
is formed leading to a spontaneous collapse of the MT network
to a ground state characterised by say a special configuration
of the displacement field $(j, m)$. This ground state
will be conformally invariant, and therefore a true vacuum
of the string, only after complete evaporation of the black hole,
which however would keep  memory of the particular collapse
mode in the `value' of the constant added to the dilaton field,
or other $W$-charges. This
reflects the existence of additional exactly marginal
deformations, consisting of global modes only,
that are not direclty accessible by local scattering experiments,
in the context of the low energy theory of propagating modes
(displacement field configurations $\psi $ in the model
of ref. \cite{mn}).
In such a case, the resulting ground state will be infinitely
degenerate, which would solve the problem of {\it memory
capacity}. Breaking of this degeneracy would correspond to
selecting a given set of expectation values for the dilaton field
ant the remaining $W$-hair charges,
which we believe corresponds to the {\it memory printing}
process, i.e. storage of information by a selection
of a given ground state. A new information would then
choose a different value of the dilaton field or other $W$-hair charges
etc. This provides a new and satisfactory mechanism of {\it memory
recall} in the following sense: if a new pulse happens to
correspond to the same set of (conserved)
$W$-hair moduli
configurations~\cite{emn}, then the associated virtual black hole
will be characterised by the same set of quantum hair, and
then the same memory state is reached {\it asymptotically}
(recall). The irreversible arrow of time, endemic in Liouville
string theory~\cite{emn} explains naturally why ``only
the past can be recalled''~\cite{vitiello}.
Indeed,
as we mentioned above, in
the presence of a space-time foamy environment,
characterised by the virtual appearence and evaporation of black holes,
there is a coupling of global modes to the propagating modes.
The environmental global
modes match in a special way with the propagating mode  $j=-\frac{1}{2}$
$m=0$, which is the
zero mode of the (massles)
tachyon corresponding to the tachyon background
of a two dimensional black hole which constitutes the {\it ground}
state or {\it memory} state of our system. This coupling
is necessary
so as to form {\it exactly}
marginal deformations~\cite{emn,chaudh}. This is a {\it special
coding} which were it not for the infinte degeneracy
of the black hole space time would lead to
a restricted memory capacity\footnote{It should
be noted at this stage that the various other
(infinite) states corresponding to continuous representations
of the $SL(2,R)$ symmetry that pertain to various
tachyon modes do not constitute memory states,
because, as mentioned earlier,
they are just {\it excitations} about the ground state.}.
However, the importance of this
coupling lies on the fact that it
leads to a time arrow in the way explained briefly above
and in detail in ref. \cite{emn}.
A fuller account
of these considerations will be given in a forthcoming
publication of ours~\cite{emncoding}.
We cannot resist in
pointing out that the
existence of such coded situations in memory cells
bears an interesting resemblence with DNA coding,
with the important difference, however, that
here it occurs in the model's state space.
\pr
At this point we would like to draw an analogy of our approach
to dissipative models for brain function, within local field theory
framework, advocated in ref. \cite{vitiello}. As observed in
ref. \cite{vitiello} the doubling of degrees of freedom
which appears necessary for a canonical quantization
of an open system~\cite{umez},
is essential in yielding~\cite{harm} a {\it non-compact}
$SU(1,1)$ symmetry for the system of damped harmonic oscillators,
used as a toy example for simulating quantum brain physics.
The quantum numbers of such a system are the $SU(1,1)$ isospin
and its third component, $j \in Z_{\frac{1}{2}}, m \ge |j|$.
The memory (ground) state corresponds to $j=0$ and there is a huge
degeneracy  characterised by the various coexisting (infinite)
eigenestates
of the Casimir operator  for the $SU(1,1)$ isospin.
The open-character of the system introduces a time arrow
which is associated with the {\it memory printing} process and
is compatible with the `observation' that
`only the past can be recalled'~\cite{vitiello}.
The crucial difference of our
string case is that the ground state of the string system, which
is conformally invariant, is actually a state with given quantum
numbers $j=-\frac{1}{2}, m=0$ of
the $SL(2,R)$ isospin in the asymptotically
flat space time case.
The degeneracy occurs, as we have already mentioned, as a result
of the `existence of an environment' of global modes, inaccessible
by local scattering operations of the brain, which lead to
exactly marginal deformations shifting the ground state
value of the dilaton field, or in general leading
to an infinity of $W$-hair charges.
\pr
The environmental entanglement of the global modes, the non-equilibrium
(in target space) nature of the temporal evolution and inevitably its
stochastic nature~\cite{emninfl},
as well as the fact that an irreversible time
arrow arises in our stringy approach~\cite{emn},
makes our non-critical string model of brain function
qualitatively
similar to the dissipative approach of ref. \cite{vitiello}.
However, in our system {\it energy} is conserved on the average, as
a result of the renormalizability of the world-sheet $\sigma$-model,
unlike the
generic dissipative system case. Thus this feature of our model
is more likely to
correspond~\cite{mn}
to realistic biological systems which are believed to
transfer energy
without dissipation~\cite{lal}.
Moreover our non-critical stringy approach is capable of
analysing real quantum gravity effects that might be responsible
for certain aspects of {\it conscious}
quantum brain function, according to the ideas
of ref. \cite{HP,dn,mn}.
Our model incorporates naturally an {\it arrow } of time,
at a {\it microscopic} level, resulting from
integrating out delocalised global (string) states that cannot
be accessible by local scattering experiments.
The associated collapse of the wave function,
as a result of quantum gravity effects, that couple
the global modes to the low-energy (observable) world, makes
a connection of this microscopic -
and thereby macroscopic and biological - time arrow, to a
{\it conscious} time arrow generated by {\it successive}
collapses in the MT cellular networks.
\pr
It should be noted, however, that
despite the appealing
features encompassed by the above ideas, very little is known
about the actual mechanisms of energy transfer in
brain cells, and therefore any claims about a
theoretical understanding of {\it conscious} perception
would be inappropriate at this stage. However, we believe that
daring of putting down some modest attempts to simulate
some apsects of the physics of the brain by capturing
(at least in a {\it qualitative} way)   certain {\it key} features
of model brain systems,
such as quantum {\it integrability} as advocated in ref. \cite{mn},
is worthy and deserves further quantitative investigations. We hope
to come back to these issues in the near future.

\pr
\nk {\bf Acknowledgements}
\pr
The work of D.V.N. is supported in part by D.O.E. grant
DE-FG05-91-ER-40633.

\end{document}